\documentclass[10pt, conference, compsocconf]{IEEEtran}

\ifCLASSINFOpdf
\else
  \usepackage[dvips]{graphicx}
\fi
\usepackage[cmex10]{amsmath}
\usepackage{amssymb}
\usepackage[tight,footnotesize]{subfigure}
\begin{document}
%
\title{Quantitative Description of Pedestrian Dynamics with a Force-based Model}
\author{\IEEEauthorblockN{Mohcine Chraibi\IEEEauthorrefmark{1},
Armin Seyfried\IEEEauthorrefmark{1},
Andreas Schadschneider\IEEEauthorrefmark{2} and
Wolfgang Mackens\IEEEauthorrefmark{3}}
\IEEEauthorblockA{\IEEEauthorrefmark{1}J\"ulich Supercomputing Centre, \\
Forschungszentrum J\"ulich GmbH,
52425 J\"ulich, Germany\\ Email: m.chraibi@fz-juelich.de, a.seyfried@fz-juelich.de}
\IEEEauthorblockA{\IEEEauthorrefmark{2}Institute of Theoretical Physics,  University of Cologne,\\ D-50937  Cologne, Germany\\
Email: as@thp.uni-koeln.de}
\IEEEauthorblockA{\IEEEauthorrefmark{3}Hamburg University of Technology,\\
 21071 Hamburg, Germany\\
Email: mackens@tuhh.de
}
}


\maketitle

\begin{abstract}
  This paper introduces a space-continuous force-based model for
  simulating pedestrian dynamics.  The main interest of this work is
  the quantitative description of pedestrian movement through a
  bottleneck. Measurements of flow and density will be presented and
  compared with empirical data.  The results of the proposed model
  show a good agreement with empirical data. Furthermore, we emphasize
  the importance of volume exclusion in force-based models.

\end{abstract}


%
\IEEEpeerreviewmaketitle

\section{Introduction}
Recently pedestrian dynamics has been gaining increasing interest.
One focus of this research area is the security of people, e.g.,
trying to optimize evacuation processes by minimizing evacuation
time. To achieve this ``benchmark'' architects and civil engineers
need to have an idea about the minimal necessary width of exit
doors, their optimal placement, length of escape routes, etc.
Usually legal regulations and other descriptive specifications in
handbooks provide requirements, which are in general not flexible
enough for complex buildings. However, simulations of pedestrian
dynamics offer the possibility to analyze and understand the dynamics
of pedestrian streams in the building to be designed. This is useful
to ease decisions e.g., dimensioning of emergency doors, and make
them more realistic and  adapted to different architectures.

In this work, we address the possibility of describing quantitatively the
pedestrian dynamics, by proposing a space-continuous model.
In section 2 the theory of mathematical modeling of pedestrian
dynamics is briefly discussed. In section 3 our model is introduced.
Then in section 4 numerical results of pedestrian flow through
bottleneck and density measurements inside and in front of the
entrance to a bottleneck will be presented. Section 5 gives concluding
remarks.

\section{Mathematical models}
There is a wide range of mathematical models to simulate pedestrian
dynamics.  Generally, these models are subdivided into macroscopic and
microscopic models~\cite{Schadschneider}. In macroscopic models the
system is described by variables like crowd-density and flows of
continua, whereas microscopic models consider the movement of
individual persons separately. Microscopic models can be subdivided
into models which are discrete or continuous in space.  Widely-used discrete
models for pedestrian dynamics are Cellular Automata models (CA),
which describe phenomena in space-time by assigning discrete states to
a grid of space-cells. These cells can be ``occupied" by a pedestrian or 
``empty". Thus the movement of pedestrians in space is done by passing 
them from cell to cell in discrete time  by simple rules that reflect
e.g.\ psychological aspects of the motion.

Space-continuous force-based models determine the continuous movement of well
distinguished individuals from their desires to reach certain targets
and from influences of the space geometry as well as actual states and
positions of the other individuals.  Inspired by the field theory of
Lewin~\cite{Lewin1951}, force-based models assume that not only
physical forces do affect the movement of pedestrians but similarly do
\textit{social} forces. 
A mathematical description of such \textit{social} forces
that determine the movement of pedestrians was first given by
Helbing and Moln\'{a}r in 1995 \cite{Helbing1995}.

Though force-based models pose quite some computational problems that
do not occur in CA, their investigation appears to be worthwhile,
since they permit higher resolution of geometry and time.

 Force-based models take Newton's second law of dynamics as 
guiding principle. The movement of each pedestrian is thus described by
\begin{equation}
\overrightarrow{F_{i}} = \sum_{i\neq j}^{\tilde N}
 \overrightarrow{F_{ij}^{rep}} + \sum_{B} \overrightarrow{F_{iB}^{rep}}
 + \overrightarrow{F_{i}^{drv}} = m_i\overrightarrow{a_i}.
\label{eq:maineq}
\end{equation}
Here $\overrightarrow{F_{ij}^{rep}}$ denotes the repulsive force
exerted by pedestrian $j$ on pedestrian $i$.
$\overrightarrow{F_{iB}^{rep}}$ is the repulsive force emerging from
borders (e.g.\ walls) and $\overrightarrow{F_{i}^{drv}}$ is a driving 
force. $m_i$ is the mass of pedestrian $i$ and $\tilde N$ is the number of
pedestrians acting on pedestrian $i$. Pedestrians try to avoid
collisions and contact with other pedestrians and objects by changing
their direction.  The repulsive forces are introduced to model this avoidance of physical contact.
They can be social forces \cite{Helbing1995,Seyfried2006,Yu2005} as
well as physical forces ~\cite{Helbing2004,Parisi2007}, where physical
forces come into play mainly to prevent overlapping with
other pedestrians and with obstacles.  
Several different forms for these repulsive forces have
been proposed. The driving force $\overrightarrow{F_{i}^{drv}}$
represents the intention of a pedestrian to  move
towards a given destination.

The set of equations~(\ref{eq:maineq}) for all pedestrians
results in a high-dimensional system of second order ordinary differential
equations. The time evolution of the positions and velocities 
of all pedestrians is then obtained by numerical integration.

Most force-based models describe the movement of pedestrians
qualitatively well: Self-organisation phenomena e.g., lane
formations~\cite{Helbing1995,Helbing2004,Yu2005}, oscillations at
bottlenecks~\cite{Helbing1995,Helbing2004}, ``faster-is-slower'' effect
~\cite{Lakoba2005,Parisi2007}, and clogging at exit
doors~\cite{Helbing2004,Yu2005} are reproduced.  These achievements
indicate that the models could reliably describe pedestrian dynamics.
However, a qualitative description is not sufficient if one has to
make reliable statements about critical processes, e.g., emergency
egress.\\ 
Moreover, rather often software implementations of the model 
do not rely on one sole approach. Especially in high density situations
the simple numerical treatment has to be supplemented by
additional techniques to obtain sensible results. Examples are
restrictions on state variables and sometimes even totally
different procedures which replace the above equations of
motion~(\ref{eq:maineq})  to avoid overlappings among pedestrians
\cite{Lakoba2005, Yu2005} and negative and high velocities
\cite{Helbing1995}.

Necessarily a model has to describe  pedestrian dynamics in a
quantitative manner, e.g., by reproducing the fundamental diagram,
the flow through bottlenecks and the density inside and in front of
the entrance of a bottleneck. Additionally, for further
investigation, it is advantageous that it uses a single set of
forces.

In the next section, we propose such a model, which is based on a
single set of formulae. Furthermore the model incorporates free
parameters which allow adaption of the model to quantitative data.

\begin{figure}
\begin{center}
\includegraphics[width=50mm,keepaspectratio]{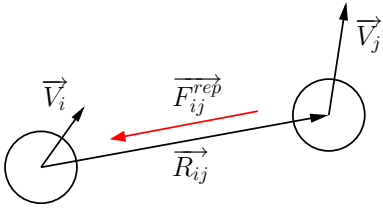}
\caption{Direction of the repulsive force acting on a pedestrian}
\label{fig:fuss}
\end{center}
\end{figure}
\section{A Modified Centrifugal Force Model}

Pedestrians are modelled as circles moving in two-dimensional space,
see Figure~\ref{fig:fuss}. We model the repulsive force exerted by
pedestrian $j$ on pedestrian $i$ as depending on the inverse of their
distance. Additionally, relative velocity between two pedestrians is
taken into account and assumed to influence the strength of the
repulsive force. Moreover, we assume proportionality between the
repulsive forces acting on a pedestrian $i$ and its desired velocity.
Thus a pedestrian tending to move with high velocity ``feels'' intense
repulsive actions.

The repulsive force, as a dimensionless relationship between the above
mentioned quantities, is defined by

 \begin{equation}
\overrightarrow{F_{ij}^{rep}}=-m_i K_{ij}\frac{(\nu \parallel
\overrightarrow{V_i^0}\parallel + V_{ij})^2}{\parallel 
\overrightarrow{R_{ij}}\parallel-\frac{1}{2}(D_i(\parallel\overrightarrow{V_i}\parallel)+D_j(\parallel\overrightarrow{V_j}\parallel)}
\overrightarrow{e_{ij}},
\label{eq:frep}
\end{equation}

where
\begin{equation}
\overrightarrow{e_{ij}} = \frac{\overrightarrow{R_{ij}}}{\parallel 
\overrightarrow{R_{ij}} \parallel},
\qquad\text{with}\quad
\overrightarrow{R_{ij}} = \overrightarrow{R_j} -\overrightarrow{R_i},
\label{eq:Re}
\end{equation}
and
\begin{equation}
 V_{ij} = \frac{1}{2}[(\overrightarrow{V_i}-\overrightarrow{V_j})\cdot
\overrightarrow{e_{ij}} + |(\overrightarrow{V_i}-\overrightarrow{V_j})
\cdot\overrightarrow{e_{ij}}|],
\label{eq:relv}
\end{equation}
and
\begin{equation}
K_{ij} = \frac{1}{2}\left[\frac{\overrightarrow{V_i}\cdot
\overrightarrow{e_{ij}}+ \mid 
\overrightarrow{V_i}\cdot
\overrightarrow{e_{ij}} \mid} {\parallel \overrightarrow{V_i}\parallel}\right].
\label{eq:K}
\end{equation}
Here $\overrightarrow{V_i^0}$ denotes the desired velocity of
pedestrian $i$ and $m_i$ its mass\footnote{In all simulations the
  mass is set to one for all pedestrians.}.  $V_{ij}$ is the
projection of the relative velocity of pedestrian $j$ and pedestrian
$i$ onto the direction of $\overrightarrow{e_{ij}}$,\, $ \parallel
\overrightarrow{R_{ij}} \parallel $ is the distance between $i$ and
$j$, $K_{ij}$ is a coefficient that reduces the action-field of the
repulsive force to the angle of vision of each pedestrian
($180^\circ$), and $\overrightarrow{e_{ij}}$ is the normalised
direction-vector between $i$ and $j$. $\parallel . \parallel$ denotes
the Euclidian norm in ${\mathbb R}^2 $ and $| . |$ the absolute value
in ${\mathbb R}$.

By means of the parameter $\nu$ the strength of the force can be
adjusted while the diameter of pedestrians
$D_i(\parallel\overrightarrow{V_i}\parallel)$ depends linearly on the
velocity. This incorporates the dynamic space
requirement of pedestrians, modeling the fact that faster pedestrians require more space than slower pedestrians, due to increasing step sizes~\cite{Seyfried2006}.

\begin{equation}
D_i= D_a + D_b\parallel\overrightarrow{V_i}\parallel
\label{eq:diam}
\end{equation}
with free parameters $D_a$\, and $D_b$.

We use the driving force as defined in~\cite{Helbing1995}:
\begin{equation}
\overrightarrow{F_{i}^{drv}} = m_i \frac{V_{i}^0 - \parallel
\overrightarrow{V_{i}}\parallel}{\tau}\overrightarrow{e_i^0},
\label{eq:fdrv}
\end{equation}
with $\overrightarrow{V_{i}}$ the velocity of pedestrian $i$,
$\overrightarrow{e_i^0}$ its desired direction and $\tau$ a time
constant.

The above model has been developed from the Centrifugal Force Model
(CFM) of Yu et al.~\cite{Yu2005}. Their expression for the
repulsive force reads as follows:
\begin{equation}
 \overrightarrow{F_{ij}^{rep}} = -m_i K_{ij} \frac{V_{ij}^2}{\parallel
\overrightarrow{R_{ij}}\parallel}\overrightarrow{e_{ij}},
\label{eq:CFMfrep}
\end{equation}
with the quantities $K_{ij}$, $V_{ij}$, $\overrightarrow{R_{ij}}$
and $\overrightarrow{e_{ij}}$ as defined in Equations~(\ref{eq:K}),
(\ref{eq:relv}), and (\ref{eq:Re}).

When pedestrians come near to one another the repulsive force is
expected to grow. In the CFM, this is in general not the case due to the
relative velocity term in the nominator of the repulsive
force~(\ref{eq:CFMfrep}). If both the distances between two
pedestrians and their relative velocities are small, their quotient
does not grow as expected. Consequently the repulsive force does not
become high enough to prohibit overlappings.  Introducing the intended
speed in the nominator of the repulsive force~(\ref{eq:frep}) we
eliminate this side-effect.  Furthermore, there is a compensation
between repulsive and driving forces at low velocities, which damps
oscillations. Due to these changes we can do in most cases without the
extra Collision Detection Technique (CDT) which takes over control in
\cite{Yu2005} in case of formation of dense crowds.

\section{Numerical results}

The solution of the initial value problem~(\ref{eq:maineq}) was done
in all simulations using a fourth-order Runge-Kutta scheme with
a fixed-step size of  $\Delta t =0.01\;$s. The repulsive interaction between two pedestrians is neglected when the distance between them is larger than two meters. 

In order to investigate the influence of our modifications,
different model approaches were simulated. Then the results were
compared  with empirical data of pedestrian flow through a
bottleneck.

The flow of $60$ pedestrians through the bottleneck situation as
described in~\cite{Seyfried2007} was simulated. The width of the
bottleneck was  changed from $0.8\;$m to $1.2\;$m in steps of
$0.1\;$m. The free parameter $\nu$ in Equation~(\ref{eq:frep}) is
set to $0.28\;\text{m/s}$ for the repulsive force among pedestrians
and to $0.4\;\text{m/s}$ for the repulsive force emerging from
obstacles.
For the free parameters in Equation~(\ref{eq:diam}) we set
\begin{equation*}
 D_a=0.2\;\text{m} \;\;\text{and}\;\; D_b=0.2\;\text{s}.
\end{equation*}

\begin{figure}[htp]
  \begin{center}
    \subfigure[Flow measurement with the modified CFM without
     CDT]{\label{fig:flowa}\includegraphics[scale=0.70]{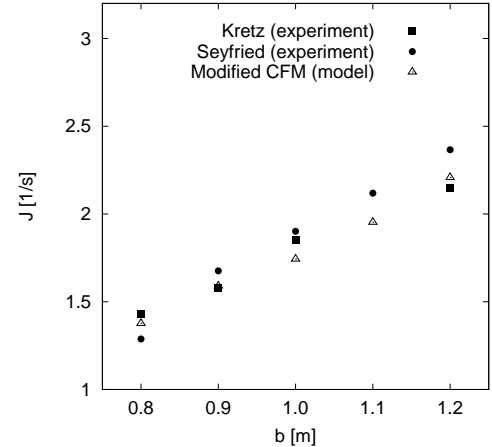}}
     \hfill
\subfigure[Flow measurement with original CFM without CDT.
Without CDT the flow values are
unrealisticly high.]{\label{fig:flowb}\includegraphics[scale=0.70]{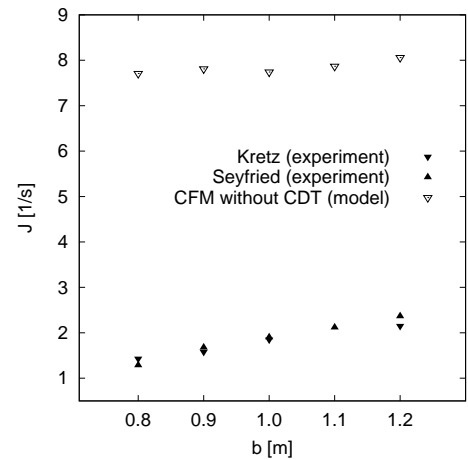}}
 \hfill
\subfigure[Flow measurement with CDT without
repulsive forces.]{\label{fig:flowc}\includegraphics[scale=0.70]{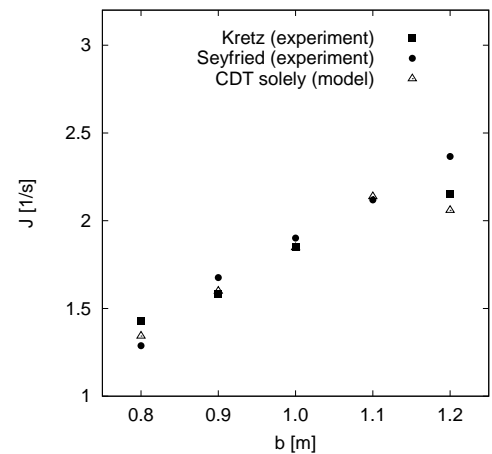}}

\end{center}
  \caption{Flow measurement with tree different models in
  comparison with empirical data  \cite{Schadschneider}.}
  \label{fig:flowm}
\end{figure}

Several parameter values were tested. With this parameter set the results of the simulations are in good agreement with the
empirical flow measurements presented in~\cite{Schadschneider}, see
Figure~\ref{fig:flowa}. For comparison the flow through the same
bottleneck was measured by the original CFM without CDT. The results
presented in Figure~\ref{fig:flowb} show very high values of the flow.
These are possible, because CDT does no more care for volume
exclusion.

Actually, control of bottleneck-flow seems to be
dominated by the CDT in the CFM-approach. To back this hypothesis we
measured pedestrian flow through the bottleneck with a CFM like model
without repulsive forces, managing collisions with CTD, only.  The
gained values are in the range of experimental data, see
Figure~\ref{fig:flowc}. Volume exclusion of pedestrians seems thus to
be a favorable feature of successful models.

\begin{figure}[htp]
\centering
\includegraphics[height=3cm, width=8cm]{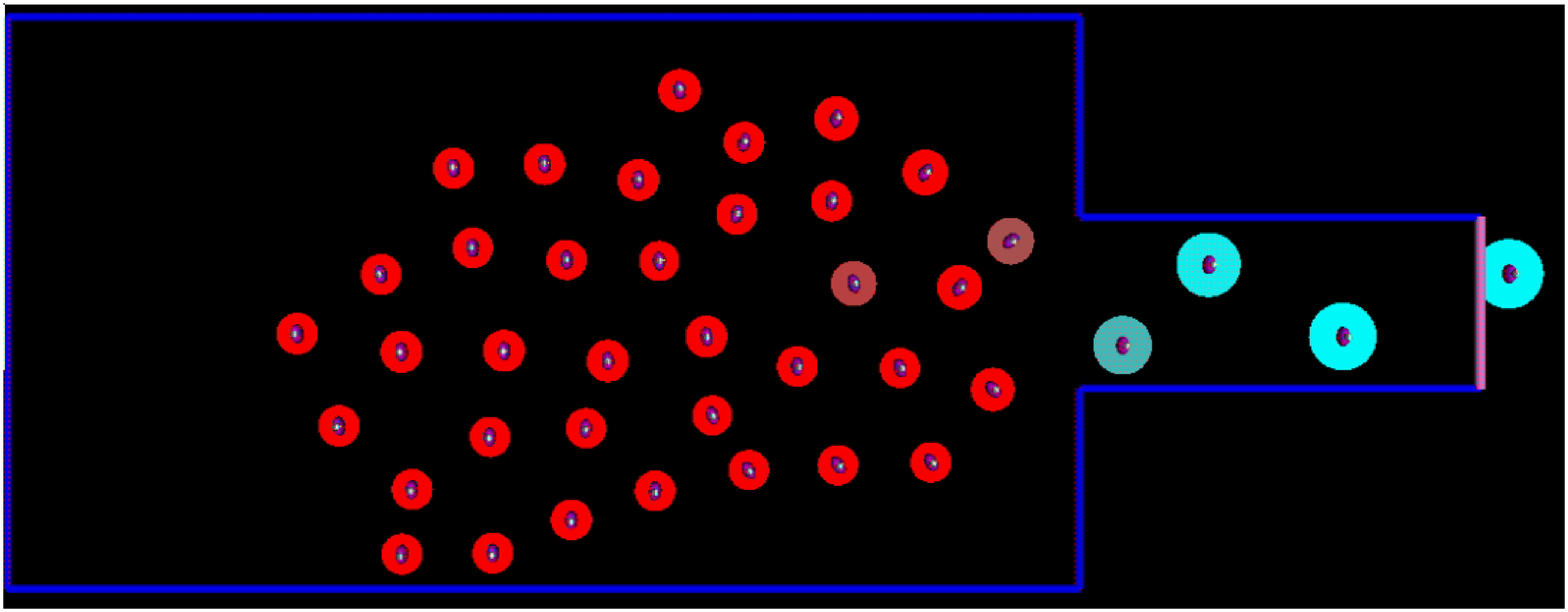}
\caption{Screenshot of a simulation: The density of pedestrians is 
  measured by counting the number of pedestrians that are located
  inside one of the measurement rectangles at each integration time.
  The area of the appropriate rectangle divides the result.}
\label{fig:scrshot}
\end{figure}

\begin{figure}[htp]
  \begin{center}
    \subfigure[Density in front of the entrance to the
    bottleneck]{\label{fig:rhovor}\includegraphics[scale=0.70]{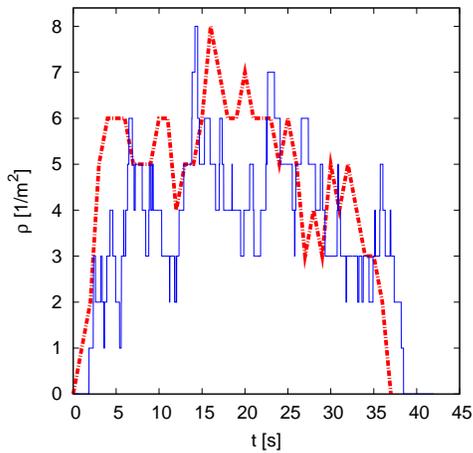}}
    \hfill \subfigure[Density inside the
    bottleneck]{\label{fig:rhoin}\includegraphics[scale=0.70]{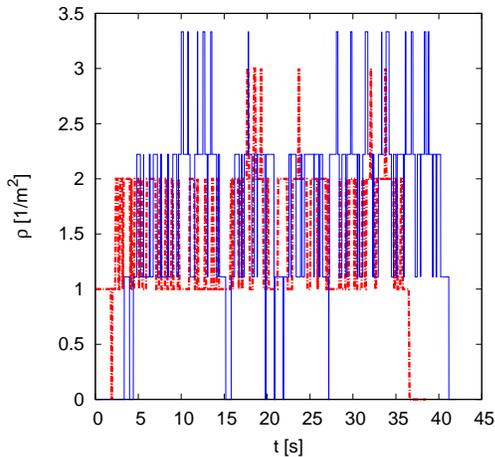}}
  \end{center}
  \caption{Density measurements: The simulation results (blue lines) are 
    in good agreement with the empirical data presented resp. in
    \cite{Seyfried2007c} and \cite{Seyfried2007}.}
  \label{fig:rho}
\end{figure}

A second validation of our modifications of CFM comes from
measurements of density inside the bottleneck as well as in front of
the entrance to the bottleneck, see Figure~\ref{fig:scrshot}. The
density in front of the entrance to the bottleneck is presented in
Figure~\ref{fig:rhovor}.  The results are in good agreement with the
experimental data in~\cite{Seyfried2007c}. Additionally, the measured
density values inside the bottleneck are in accordance with the
published empirical results in~\cite{Seyfried2007}, see
Figure~\ref{fig:rhoin}.

Though the model is already a useful tool to describe quantitatively
the bottleneck-situations as defined in~\cite{Seyfried2007}, it is
not yet fully satisfactory. Depending on the specific situation
oscillations and overlappings can still occur. Determination of
uniformly usable parameters is a difficult endeavour; cf.
      the parameter estimation in SimWalk~\cite{Steiner2007}.

\section{Conclusion}

The proposed space-continuous force-based model quantitatively
describes the movement of pedestrians in 2D-space in an almost
satisfactory way.  Besides being a remedy for numerical instabilities
in CFM the model simplifies the approach of Yu et al. since we can
dispense with their extra CDT without deteriorating performance.  The
implementation of the model is straightforward and does not use any
restrictions on the velocity as was the case e.g.,
in~\cite{Helbing1995}.

 Simulation results show good agreement with empirical data.
The model contains natural free parameters which can further be
tuned to adapt the model to certain scenarios. Introducing
density-dependence of the strength of the repulsive force promises
further ameliorations of the model.

 Furthermore we have provided some evidence that
volume exclusion is an important
issue in describing pedestrian dynamics.




%


\end{document}